\documentclass[preprint]{imsart}
\usepackage{amsmath}
\usepackage{amssymb, epsfig, natbib}
\renewcommand{\Pr}{\bm{P}}
\newcommand{\mat}[1]{\text{\mbox{\boldmath$#1$}}}
\newcommand{\bm}[1]{\text{\mbox{\boldmath$#1$}}}
\newtheorem{example}{Example}
\newtheorem{definition}{Definition}
\newtheorem{theorem}{Theorem}
\begin{document}
\begin{frontmatter}
\title{Multiple Test Functions and Adjusted p-Values for Test Statistics with Discrete Distributions}
\runtitle{Multiple Test Functions}
\runauthor{J. D. Habiger}
\author{Joshua D. Habiger}
\address{Oklahoma State University, Department of Statistics, 301-G MSCS, Stillwater, OK, 74078-1056, USA.  email: jhabige@okstate.edu}
\begin{abstract}
The randomized $p$-value, (nonrandomized) mid-$p$-value and abstract randomized $p$-value have all been recommended for testing a null hypothesis whenever the test statistic has a discrete distribution.   This paper provides a unifying framework for these approaches and extends it to the multiple testing setting.  In particular, multiplicity adjusted versions of the aforementioned $p$-values and multiple test functions are developed.  It is demonstrated that, whenever the usual nonrandomized and randomized decisions to reject or retain the null hypothesis may differ, the (adjusted) abstract randomized $p$-value and test function should be reported, especially when the number of tests is large.  It is shown that the proposed approach dominates the traditional randomized and nonrandomized approaches in terms of bias and variability.  Tools for plotting adjusted abstract randomized $p$-values and for computing multiple test functions are developed. Examples are used to illustrate the method and to motivate a new type of multiplicity adjusted mid-$p$-value.
\end{abstract}
\begin{keyword}
abstract randomized $p$-value;
adjusted $p$-value;
decision function;
fuzzy $p$-value;
mid-$p$-value;
randomized $p$-value;
test function;
\end{keyword}
\end{frontmatter}
\baselineskip = 22pt
\section{Introduction}
High throughput technology now routinely yields large and complex data sets which have forced statisticians to rethink traditional statistical inference. For example, a basic microarray data set allows for several thousand null hypotheses to be tested simultaneously, and it is now widely accepted that multiplicity effects cannot be ignored.  Consequently, dozens, if not hundreds, of multiple hypothesis testing procedures have been developed in recent years; see \cite{Fer08} or the books by \cite{WesYou93, DudVan08, Efr10} for reviews.  Ultimately, which of these multiple testing procedures is employed will depend on the data structure, the cost of falsely rejecting or failing to reject null hypotheses, and even taste.  In this paper, we are interested in the setting when each test statistic has a discrete distribution, which arises when data are dichotomous in nature or when utilizing nonparametric rank-based test statistics.

Challenges arising when test statistics have discrete distributions are evident even when testing a single null hypothesis.  To illustrate, consider the following example (see \cite{Cox74} for more details):
\begin{example} \label{example}
Suppose that $X$ is a binomial random variable with $n = 10$ and success probability $p$, and that the goal is to test null hypothesis $H_0:p = 1/2$ vs. alternative hypothesis $H_1:p > 1/2$.  Define size-$0.05$ test function
$$\phi(x) = \left\{\begin{array}{lr}
1 & x>8\\
0.89 & x = 8 \\
0 & x<8\end{array}\right..\label{example}
$$
\end{example}
In Example \ref{example}, $\phi(x)$ is said to have size 0.05 because its expectation is equal to $0.05$ under $H_0$. To implement $\phi(x)$, $H_0$ is rejected if $\phi(x) = 1$ and is not rejected, or retained, if $\phi(x) = 0$.  If $\phi(x) = 0.89$, several different approaches could be taken.

Earlier authors \citep{NeyPea33, Pea50, Toc50, Pra61} proposed a randomized testing strategy, which is implemented by generating $U=u$ from a uniform$(0,1)$ distribution and rejecting $H_0$ if $u\leq\phi(x)$, or equivalently by rejecting $H_0$ if the randomized $p$-value defined by
$$p(x,u) = \Pr_{X}^0(X>x) + u\Pr_{X}^0(X=x),$$
where $\Pr_{X}^0(\cdot)$ denotes a probability computed under $H_0$, is less than or equal to $0.05.$  This randomized decision rule ensures that the probability of erroneously rejecting $H_0$ is indeed 0.05 but is sometimes regarded as impractical because the final decision could depend on the value of the independently generated $u$.  See \cite{HabPen11} for a discussion.  Another strategy is to reject $H_0$ if a nonrandomized $p$-value, such as the mid-$p$-value in \cite{Lan61} defined by $p(x,1/2)$ or the natural $p$-value defined by $p(x,1)$, for example, is less than or equal to 0.05.  See \cite{Agr92, Agr01, Yan04, Agr07} for other examples.  The advantage of this strategy is that it does not make use of an independently generated random variable.  However, neither of these nonrandomized decision rules have size 0.05.   In an effort to avoid the mathematical disadvantages of nonrandomized $p$-values and at the same time avoid generating a uniform$(0,1)$ variate, \cite{GeyMee05} proposed the fuzzy, also sometimes called the abstract randomized, $p$-value, which can be defined as $p(x,U)$, for example.  Here, we write capital $U$ to denote an (unrealized) uniform(0,1) random variable.  Note that from a probability perspective, $p(x,U)$ also represents an unrealized random variable. \footnote{We will not be concerned with the distinction between abstract randomized $p$-values and fuzzy $p$-values because, as mentioned in \cite{GeyMee05}, they are ``different interpretations of the same mathematical object.''}  Most detailed studies of these three approaches (see, for example, \cite{Bar89, GeyMee05}) have focused on recommending one of the aforementioned $p$-values.  This is not our goal here.

One goal of this paper is to illustrate that the test function and the abstract randomized $p$-value provide information that is likely of interest in settings when the randomized and nonrandomized $p$-values need not yield the same final decision.  To see why, let us consider a second binomial example.
\begin{example} \label{example2}
Consider the same null and alternative hypotheses as in Example \ref{example}, but suppose that $n = 11$.  The size-$0.05$ test function is defined
$$\phi(x) = \left\{\begin{array}{lr}
1 & x>8\\
0.21 & x = 8 \\
0 & x<8\end{array}\right..
$$
\end{example}
If $X>8$ or if $X<8$ then the final decision is clear in both Example \ref{example} and \ref{example2} whether opting for a randomized or nonrandomized decision rule.  However, if $X = 8$ then the randomized decision depends on the value of the generated $u$ in both examples.  On the other hand, the mid-$p$-value for Example \ref{example} is $p(8,1/2) = 0.033$ and results in a rejected null hypothesis while the mid-$p$-value for Example \ref{example2} is $p(8,1/2) = 0.073$ and results in a retained null hypothesis.  While the practitioner will likely be satisfied with the nonrandomized decision to reject $H_0$, s/he may feel uneasy about the nonrandomized decision to retain $H_0$ because the $p$-value is only slightly larger than 0.05.  We may try to appease the practitioner by explaining that the size of our nonrandomized decision rule was only approximately $0.05$.  However, a more informative statement is available.  In particular, it can be verified that the size of the mid-$p$-value-based decision rule in Example \ref{example2} is $\Pr_X^0(X>8) = 0.033$. Hence, we may additionally report that our decision to retain $H_0$ is conservative.  As it turns out, this conservative behavior can be traced back to the fact that $\phi(8) = 0.21 \leq 1/2$.  Likewise, the mid-$p$-value-based decision rule in Example \ref{example} is liberal due to the fact that $\phi(8) = 0.89>1/2$.  Thus, the test function provides valuable information in this setting.  This notion is formalized in Section 2 and motivates the proposed hybrid-type hypothesis testing procedure, where $\phi(x)$ and its corresponding abstract randomized $p$-value are reported in addition to the usual randomized or nonrandomized $p$-values/decisions whenever discrepancies between decisions could exist, i.e. whenever different values of $u$ could lead to different decisions.

Of course, as we saw in the above examples, discrepancies need not occur on a single run of the experiment, in which case the choice to opt for a randomized, mid- or natural $p$-value does not affect the final decision.  However, when testing many null hypotheses simultaneously such discrepancies are likely to occur more than once and the manner in which they are handled can have a significant impact on the final analysis. For example, in the analysis of a microarray data set in Section 4, 153 additional null hypotheses are rejected when utilizing the usual mid-$p$-values rather than natural $p$-values.  On the other hand, randomized $p$-values and the mid-$p$-values proposed in Section 5 lead to some, but not all, of the 153 null hypotheses being rejected.  As in the single null hypothesis testing case, the \textit{multiple test function} and multiplicity \textit{adjusted abstract randomized $p$-values}, developed and studied in Section 3, can be used to describe the operating characteristics of each approach and should be reported whenever discrepancies could exist.

This paper proceeds as follows. Section 2 provides a general unifying framework for nonrandomized, randomized and abstract randomized $p$-values and their corresponding test functions. Additionally, the proposed method is introduced and analytically compared to the more standard randomized and nonrandomized approaches. Section 3 develops a computationally friendly and general definition of multiple test functions and their corresponding multiplicity adjusted abstract randomized $p$-values.  It also extends the proposed method into the multiple testing setting and provides an analytical assessment.  Examples in Section 4 are used to illustrate the method and to motivate a new type of adjusted mid-$p$-value in Section 5.  It is shown that the proposed adjusted mid-$p$-value is mathematically more tractable than other adjusted nonrandomized or randomized $p$-values. Concluding remarks are in Section 6 and proofs are in the Appendix.

\section{Inference for a single null hypothesis}
\subsection{$p$-Values, decision functions and test functions}
Before defining the method, we first define all of its relevant components and examine key relationships between decision functions, $p$-values and test functions.  Let $X$ be a discrete random variable with countable support $\mathcal{X}$ and let $U$ be an independently generated uniform$(0,1)$ random variable.  Assume that a null hypothesis $H_0$ specifies a distribution for $X$ and that the basic goal is to decide to reject or retain $H_0$.  As in the Introduction, denote expectations and probabilities computed under $H_0$ with respect to $X$ by $E_X^0(\cdot)$ and $\Pr_X^0(\cdot)$, respectively.  Expectations and probabilities computed with respect to $X$, but not necessarily under $H_0$, are written $E_{X}(\cdot)$ and $\Pr_X(\cdot)$.  Additionally, expectations and probabilities taken with respect $U$ are written $E_U(\cdot)$ and $\Pr_U(\cdot)$, respectively, and expectations and probabilities taken taken over $(X,U)$ are written $E_{(X,U)}(\cdot)$ and $\Pr_{(X,U)}(\cdot)$, respectively.

The main mathematical object for testing $H_0$ is a test function $\phi:\mathcal{X}\times[0,1]\rightarrow [0,1]$, written $\phi(x;\alpha)$.  In this section, the parameter $\alpha$ refers to the size of the test.  That is,  $E_X^0[\phi(X;\alpha)] = \alpha$.  We use the terminology ``size'' as opposed to ``level'' to emphasize that the aforementioned expectation is equal to $\alpha$, whereas the terminology ``level'' sometimes only indicates that the expectation is less than or equal to $\alpha$.  Assume that $\phi(x;\alpha)$ is nondecreasing and right continuous in $\alpha$ for every $x\in\mathcal{X}$ throughout this paper.  To make matters more concrete we will sometimes consider the following specific type of test function:
\begin{definition}
Let $X$ be a test statistic defined such that large values of $X$ are evidence against $H_0$.  Define size-$\alpha$ test function
\begin{equation}\phi^*(x;\alpha) = \left\{\begin{array}{lr}
1 & x>k(\alpha)\\
\gamma(\alpha) & x = k(\alpha) \\
0 & x<k(\alpha)\end{array}\right.\label{example u}
\end{equation}
where $k(\alpha)\in\mathcal{X}$ and $\gamma(\alpha)\in(0,1)$ solve $\Pr_X^0(X>k) + \gamma \Pr_X^0(X=k) = \alpha$.
\end{definition}

Now, if $\phi(x;\alpha) = 1 (0)$ then $H_0$ will be automatically rejected (retained).  When $\phi(x;\alpha)\in (0,1)$ we may force a reject or fail to reject decision via the decision function $\delta:\mathcal{X}\times[0,1]\times[0,1]\mapsto \{0,1\}$ defined $\delta(x,u;\alpha) = I(u\leq \phi(x;\alpha))$, where $I(\cdot)$ is the indicator function and $\delta(x,u;\alpha) = 1 (0)$ means that $H_0$ is rejected (retained).
Any decision function has a corresponding $p$-value statistic, which we define as in \cite{HabPen11} and \cite{Pen11}:
$$
p(x,u) = \inf\{\alpha: \delta(x,u;\alpha) = 1\}.
$$
In words, $p(x,u)$ is the smallest value of $\alpha$ that allows for $H_0$ to be rejected.  Hence, if $\alpha$ corresponds to the size of $\delta(x,u;\alpha)$ then $p(x,u)$ is the smallest size allowing for $H_0$ to be rejected.

Observe that for $\phi^*(x;\alpha)$ defined as in (\ref{example u}) and $\delta^*(x,u;\alpha) = I(u\leq \phi^*(x;\alpha))$, we recover $p$-value
\begin{equation}\label{example pu}
p^*(x,u) = \inf\{\alpha:\delta^*(x,u:\alpha) = 1\} = \Pr_{X}^{0}(X>x) + u \Pr_X^{0}(X=x).
\end{equation}
The key steps in verifying the second equality are to first write the constraint $E_X^{0}[\phi^*(X;\alpha)] = \alpha$ as $\alpha = \Pr_X^0(X>k(\alpha)) +\Pr_U(U\leq\gamma(\alpha))\Pr_X^0(X=k(\alpha))$ and then verify that
$\delta^*(x,u;\alpha) = 1$ if and only if $\alpha\geq \Pr_X^{0}(X>x) + u \Pr_X^{0}(X = x)$.  See \cite{HabPen11} for more details and examples.

In general, when $u$ is generated, we refer to $p(x,u)$ as a \textbf{randomized $p$-value} and refer to $\delta(x,u;\alpha)$ as a \textbf{randomized decision function}. When $u$ is specified, say $u = 1/2$ or $u = 1$, we refer to $p(x,u)$ as a \textbf{nonrandomized $p$-value} and $\delta(x,u;\alpha)$ as a \textbf{nonrandomized decision function}.  Finally, when $U$ is a random variable that has not yet been generated we refer to $p(x,U)$ as an \textbf{abstract randomized $p$-value} or \textbf{fuzzy $p$-value}.  Note that the mid-$p$-value in \cite{Lan61} is recovered via $p^*(x,1/2)$.  In this paper, however, we refer to any $p$-value computed via $p(x,1/2)$ as a mid-$p$-value.

Theorem 1 in \cite{Hab12} indicates that if the ultimate goal is to make a reject $H_0$ or retain $H_0$ decision, we may use either a $p$-value or its corresponding decision function.  That is,
$$
\delta(X,U;\alpha) = I(p(X,U)\leq \alpha)
$$
with probability 1.  A similar relationship between the abstract randomized $p$-value and its corresponding test function exists and is formally described below.
\begin{theorem} \label{theorem 1}
Let $U$ be uniformly distributed over $(0,1)$ and independent of $X$.  Then, for every $x\in\mathcal{X}$, $\Pr_U(p(x,U)\leq\alpha) = \phi(x;\alpha)$.
\end{theorem}

Theorem \ref{theorem 1}  states that reporting the value of a test function is (mathematically) akin to reporting the proportion of the distribution of $p(x,U)$ that is below $\alpha$.  To better understand the implications, reconsider Example \ref{example}.  Observe that if $X>8$ then $\phi(x;0.05) = 1$ and Theorem \ref{theorem 1} implies that $p(x,u)$ is less than or equal to 0.05 regardless of the value of $u$.  That is, the entire distribution of $p(x,U)$ is less than or equal to 0.05.   Likewise, if $X< 8$ so that $\phi(x;0.05) = 0$ then $p(x,U)>0.05$ with probability 1.  If $X = 8$, then $p(8,U)$ is a random variable satisfying $\Pr_U(p(8,U)\leq 0.05)= 0.89$.  Of course, in this case, we could have also verified that $p(8,U) = 0.011 + U 0.044$, i.e. that $p(8,U)$ is uniformly distributed over (0.011, 0.055), and directly verified that $\Pr_U(p(8,U)\leq 0.05) = 0.89$.  However, as we will see in Section 4, the distribution of the abstract randomized $p$-value can be difficult to derive in closed form.


\subsection{The method}
We are now in position to introduce the method.  Assume that $\phi(x;\alpha)$ is defined and $X=x$ is observed.
\begin{enumerate}
\item[1.]\textit{Compute $\phi(x;\alpha)$.  If $\phi(x;\alpha) = 1(0)$ then reject(retain) $H_0$ and quit.  Otherwise report $p(x,U)$ and $\phi(x;\alpha)$ and go to 2a. or 2b.
\item[2a.] Generate $U=u$, an independent uniform$(0,1)$ variate, and compute $\delta(x,u;\alpha)$ and $p(x,u)$.  If $\delta(x,u;\alpha) = 1(0)$ or if $p(x,u)\leq(>)\alpha$ then reject(retain) $H_0$.
\item[2b.] Specify $u$ and compute $\delta(x,u;\alpha)$ and $p(x,u)$.  If $\delta(x,u;\alpha) = 1(0)$ or if $p(x,u)\leq(>)\alpha$ then reject(retain) $H_0$.}
\end{enumerate}

There are two important points to be made.  First, most of the time a reject or retain $H_0$ decision is automatically made ($\Pr_X^0(X\neq 8) = 0.956$ in Example \ref{example}) and any controversy over the abstract randomized, randomized or nonrandomized $p$-value is avoided. On the other hand, if $\phi(x;\alpha)$ is not 0 or 1 then the method still reports one of the usual well-understood randomized (Step 2a) or nonrandomized (Step 2b) $p$-values/decisions, but additionally brings to forefront that the final decision rested on the value of $u$.  That is,  the method alerts the user that a discrepancy between randomized and nonrandomized decisions may exist.  Tools for characterizing the nature of the potential discrepancy are provided next.

\subsection{Assessment}
In the following Theorem and discussion, we more precisely characterize the additional information contained in $\phi(x;\alpha)$ by comparing the distribution of $\phi(X;\alpha)$ to the distribution of the randomized decision function $\delta(X,U;\alpha)$ and the nonrandomized decision function $\delta(X,u;\alpha)$.  That is, the following Theorem compares the proposed method to the usual randomized and nonrandomized methods, which skip Step 1 and automatically go to Step 2a or Step 2b, respectively.
\begin{theorem} \label{theorem 2} Let $U$ be uniformly distributed over (0,1) and independent of $X$.  Then
\begin{enumerate}
\item[C1:] $E_{U}[\delta(x,U;\alpha)] = \phi(x;\alpha)$ for every $x\in\mathcal{X}$ and hence $E_{(X,U)}[\delta(X,U;\alpha)] = E_X[\phi(X;\alpha)]$,
\item[C2:] $Var(\delta(X,U;\alpha))\geq Var(\phi(X;\alpha)).$
\end{enumerate}
For any fixed or specified value of $u$,
\begin{enumerate}
\item[C3:]  $E_X[\delta^*(X,u;\alpha)]$ is nonincreasing in $u$ and $E_X[\delta^*(X,u;\alpha)]\neq E_X[\phi^*(X;\alpha)]$ for every $\gamma(\alpha)\in(0,1)$.  In particular, $E_X[\delta^*(X,u;\alpha)] >(<) E_X[\phi^*(X;\alpha)]$ for $u\leq (>)\gamma(\alpha)$.
\end{enumerate}
\end{theorem}

In the ensuing discussion, we adopt the viewpoint that  $\delta(x,u;\alpha)$ is an estimate of $\phi(x;\alpha)$ based on data $u$.  For a similar viewpoint, see \cite{BlySta95}, where the goal was to estimate $I(H_0 \mbox{ true})$. Now, a careful inspection of Theorem \ref{theorem 2} indicates that information is lost in only reporting $\delta(x,u;\alpha)$, whether $u$ is generated or specified. From Claim C1 we recover the well-known result that the randomized decision rule is unbiased; its expectation is equal to the expectation of $\phi(X;\alpha)$.  Hence, if $\phi(X;\alpha)$ is a size-$\alpha$ test then so is $\delta(X,U;\alpha)$.  However, Claim C2 indicates that the randomized decision function is more variable than $\phi(X;\alpha)$, which reflects the fact that information is lost in reporting $I(u\leq\phi(x;\alpha))$ alone.

Claim C3 illustrates that the information lost due to the specification of $u$ is manifested as bias, in that the size of $\delta(X,u;\alpha)$ is no longer equal to $\alpha$, and provides additional information regarding the nature and degree of the bias.  First, observe that the procedure is liberal in that it rejects $H_0$ with too high of a probability if $u\leq \gamma(\alpha)$, and is conservative otherwise.  \cite{Lan61} demonstrated that the degree of this bias was typically minimized by the choice $u = 1/2$.  Equation (\ref{notequal}) in the proof of Theorem \ref{theorem 2} (see the Appendix) sheds light on this phenomon.  Here, we see that bias is attributable to the fact that $|I(u\leq\gamma(\alpha)) - \gamma(\alpha)|>0$, a quantity whose supremum over $\gamma(\alpha)$ is minimized by choosing $u = 1/2$.  Additionally, observe that if $\gamma(\alpha)$ is near 0 or 1 and $u = 1/2$, then the bias quantity $|I(u\leq \gamma(\alpha)) - \gamma(\alpha)|$ is near 0, while if $\gamma(\alpha)$ is near $1/2$, then the bias quantity is near $1/2$.  Hence, the value of $\gamma(\alpha)$ indicates whether the procedure is liberal or conservative, and by how much.  In short, we can do better than simply claiming: ``the decision rule is roughly of size-$\alpha$.''  We can additionally report, for example, that our final decision rule was ``slightly liberal'' or ``moderately conservative'',  etc.

\section{Inference for multiple null hypotheses}
The basic methodology is akin to the method in the previous section.  However, we first illustrate that care must be taken in defining multiple test functions (MTFs) and their corresponding adjusted abstract randomized $p$-values, especially when utilizing a step-wise multiple testing procedure.
\subsection{Multiple testing procedures}
Let $\mat{X} = (X_1, X_2, ..., X_M)$ be a collection of test statistics with countable support $\mathcal{X} = \mathcal{X}_1\times\mathcal{X}_2\times...\times\mathcal{X}_M$ and let $\mat{U} = (U_1, U_2, ..., U_M)$ be a collection of $M$ independent and identically distributed uniform(0,1) random variables.  Further assume $\mat{X}$ and $\mat{U}$ are independent.  Suppose that the goal is to decide whether to reject or retain the null hypothesis $H_{0m}$, which specifies a distribution for $X_m$, for each $m \in \{1, 2, ..., M\}$.

Many multiple testing procedures are defined in terms of a collection of $p$-values, which we will assume can be computed $\mat{p} = \mat{p}(\mat{x},\mat{u}) = [p_1(x_1,u_1), p_2(x_2,u_2), ..., p_M(x_M,u_M)]$, where $p_m(x_m,u_m)$ is the $p$-value for testing $H_{0m}$ with ($x_m, u_m$) defined as in the previous section.  Formally, a multiple testing procedure (MTP) is a multiple decision function  $\bm{\delta}:\mathcal{X}\times[0,1]^{M}\times[0,1]\mapsto \{0,1\}^M$ defined via $\bm{\delta}(\mat{x},\mat{u};\alpha) = [\delta_1(\mat{x},\mat{u};\alpha), \delta_2(\mat{x},\mat{u};\alpha), ..., \delta_M(\mat{x},\mat{u};\alpha)]$ where each $\delta_m$ depends on potentially all of the data $\mat{x}$ and a generated or specified $\mat{u}$ through the $p$-values $\mat{p}(\mat{x},\mat{u})$ and a ``thresholding'' parameter $\alpha$.  Again, $\delta_m(\mat{x},\mat{u};\alpha) = 0 (1)$ means that $H_{0m}$ is retained (rejected).  Here, $\alpha$ will typically correspond to the global error rate of interest, such as the False Discovery Rate (FDR) or Family-Wise Error Rate (FWER), defined $FWER = \Pr(V\geq 1)$ and $FDR = E[V/\max\{R,1\}]$ \citep{BenHoc95}, respectively, where $V$ is the number of erroneously rejected null hypotheses, also called false discoveries, and $R$ is the number of rejected null hypotheses or discoveries, and where expectations and probabilities are taken with respect to the true unknown distribution of \mat{X} and $\mat{U}$ (if $\mat{U}$ is random).  See \cite{Sar07} for other error rates.




For example, let us consider the single-step Bonferroni procedure and one of its step-wise counterparts, the \cite{Hol79} procedure.  For a more general definition of step-wise MTPs see \cite{Tam98} or see \cite{Fer08} for a review.  The Bonferroni procedure rejects $H_{0m}$ if $p_m(x_m,u_m)\leq \alpha/M$, i.e. is defined $\delta_m^{Bon}(x_m,u_m;\alpha) = I(p_m(x_m,u_m)\leq \alpha/M)$ for each $m$.  The Holm procedure makes use of the ordered $p$-values $p_{(1)} \leq p_{(2)} \leq ... \leq p_{(M)}$.  Specifically, it defines
\begin{equation}\label{Hol ineq}
k(\mat{p}) = \max\left\{m: p_{(j)}\leq \alpha \frac{1}{M-j+1} \mbox{ for all }j\leq m\right\}
\end{equation}
and rejects the k null hypotheses corresponding to $p_{(1)}, p_{(2)}, ..., p_{(k)}$, or equivalently, defines the $m$th decision function by  $$\delta_m^{Hol}(\mat{x},\mat{u};\alpha) = I\left(p_m(x_m,u_m) \leq \frac{\alpha}{M-k(\mat{p})+1}\right).$$
Now, to see why multiple test function corresponding to multiple decision functions are not necessarily well-defined, consider defining a multiple test function corresponding the Holm MTP.  Immediately, we see that $k(\mat{p})$ is only well defined for randomized or nonrandomized $p$-values where $p$-values can be ranked and the inequality in (\ref{Hol ineq}) is well defined.

In general, step-wise procedures will make use of the ordered $p$-values and the above issue will arise.  This is problematic because step-wise procedures tend to dominate their single step counterparts \citep{Leh05}.  We should remark that \cite{KulLew09} were able to extend the (step-wise) \cite{BenHoc95} procedure to handle abstract randomized $p$-values.  In the next subsection we expand upon their idea by providing a general and computationally friendly definition of a multiple test function.

\subsection{Multiple test functions}

The idea is as follows. First, note that once a value for each $u_m$ is either generated or specified, then $p$-values can be ranked and most multiple testing procedures can be applied.  Further, recall from Claim C1, $\phi(x;\alpha) = E_U\delta(x,U;\alpha) = \int_{[0,1]}\delta(x,u;\alpha)du$.  This leads to the following definition for an MTF.
\begin{definition} \label{mtf}
Let $\bm{\delta}(\mat{x},\mat{u};\alpha)$ be a multiple testing procedure defined
$$\bm{\delta}(\mat{x},\mat{u};\alpha) = [\delta_1(\mat{x,u};\alpha), \delta_2(\mat{x,u};\alpha), ..., \delta_M(\mat{x},\mat{u};\alpha)].$$ Define multiple test function $\bm{\phi}:\mathcal{X}\times [0,1]^M\mapsto[0,1]^{M}$ by
$$\bm{\phi}(\mat{x};\alpha) = [\phi_1(\mat{x};\alpha), \phi_2(\mat{x};\alpha), ..., \phi_M(\mat{x};\alpha)]$$ where
\begin{equation*}
\phi_m(\mat{x};\alpha) = E_{\mat{U}}[\delta_m(\mat{x},\mat{U};\alpha)] = \int_0^1\int_0^1...\int_0^1\delta_m(\mat{x},\mat{u};\alpha)du_1du_2...du_M \label{PHI}
\end{equation*}
for each $m$.
\end{definition}
It should be noted that the above definition implicitly assumes that the $U_m$s are independently generated.  While it is possible to relax this condition in our definition, the resulting MTF is less tractable mathematically and in practice.  A more detailed discussion is provided in the Concluding Remarks section.  We shall additionally assume that $\phi_m(\mat{x};\alpha)$ is nondecreasing and right continuous in $\alpha$ for every $\mat{x}\in\bm{\mathcal{X}}$, which is satisfied, for example, by definition if each $\delta_m(\mat{x},\mat{u};\alpha)$ is nondecreasing and right continuous in $\alpha$ with probability 1.

A basic Monte-Carlo algorithm can be used to compute $\bm{\phi}(\mat{x};\alpha)$ and is illustrated as follows.  Of course, other integration techniques could be employed.  See, for example, \cite{RobCas04}.
\begin{enumerate}
\item For $b = 1, 2, ..., B$ generate an independent vector of $M$ i.i.d. uniform$(0,1)$ variates $\mat{U}^b =\mat{u}^b$ and compute $p_m(x_m,u_m^{b})$ for each $m$.
\item Apply the MTP to $p_1(\mat{x},\mat{u}^{b}), p_2(\mat{x},\mat{u}^{b}), ..., p_{M}(\mat{x},\mat{u}^{b})$ and record $\delta_m(\mat{x},\mat{u}^{b};\alpha) = 1(0)$ for each $m$ and each $b$.
\item Compute test function $\phi_m(\mat{x};\alpha) = B^{-1}\sum_{b=1}^{B}\delta_m(\mat{x},\mat{u}^{b};\alpha)$ for each $m$.
\end{enumerate}

To illustrate, suppose that $X_m$ has a binomial distribution with sample size $n = 10$ and success probability $p_m$, and suppose that we wish to compute the Holm MTF with $\alpha = 0.05$ for testing null hypothesis $H_{0m}:p_m=1/2$ against alternative hypothesis $H_{1m}:p_m>1/2$ for $m = 1, 2, ..., 5$.  Further, assume $\mat{X} = (10, 9, 8, 6, 5)$.  Then for $b = 1, 2, ..., 1000$ we generate $\mat{U}^b = \mat{u}^b$ as above, compute the randomized $p$-value $p_m^*(x_m,u_m^b)$ as in (\ref{example pu}) for each $m$, and compute $\delta_m^{Hol}(\mat{x},\mat{u}^b;\alpha)$.  Finally, compute $\phi_m^{Hol}(\bm{x};0.05) = B^{-1}\sum_{b=1}^B\delta_m^{Hol}(\mat{x},\mat{u}^b;\alpha)$ for each $m$.  Results are summarized in Table \ref{illustrate}.  Observe that $\delta_m^{Hol}(\mat{x},\mat{u}^b)$ was equal to 1 for every $b$ when $m = 1, 2$ and was equal to 0 for every $b$ when $m = 4, 5$.  However, 13\% of the $\delta_3^{Hol}(\mat{x},\mat{u}^b;\alpha)$s were equal to 1.  Hence, $\phi_3^{Hol}(\mat{x};\alpha) = 0.13$.
\begin{table}[h!]\begin{center}
\caption{An illustration of the Monte-Carlo integration method for computing $\bm{\phi}(\mat{x};\alpha)$.\label{illustrate} A ``$^*$'' indicates that the null hypothesis was rejected by the Holm MTP, which was applied at $\alpha = 0.05$. }
\begin{tabular}{|cccccc|}\hline
b & $p_1^*(10,u_1^b)$ & $p_2^*(9,u_2^b)$ & $p_3^*(8,u_3^b)$ & $p_4^*(6,u_4^b)$ & $p_5^*(5,u_5^b)$ \\ \hline
1 & 0.001$^*$ & 0.004$^*$ & 0.016$^*$ & 0.348 & 0.492\\
2 & 0.001$^*$ & 0.001$^*$ & 0.034& 0.363 & 0.459 \\
3 & 0.000$^*$ & 0.009$^*$ & 0.030 & 0.343 & 0.496 \\
\vdots & \vdots&\vdots & \vdots & \vdots & \vdots \\
1000 & 0.000$^*$&0.004$^*$ &0.041 & 0.355 & 0.444 \\ \hline
$\phi_m^{Hol}(\mat{x};.05)$ & 1$^*$ & 1$^*$ & 0.13 & 0 & 0 \\ \hline
\end{tabular}
\end{center}
\end{table}

\subsection{Adjusted $p$-values}
The basic idea of an adjusted $p$-value was formalized in \cite{Wri92}. Recall that the usual $p$-value is the smallest value of $\alpha$ allowing for the null hypothesis $H_0$ to be rejected by $\delta(x,u;\alpha)$.  The adjusted $p$-value for $\delta_m(\mat{x},\mat{u};\alpha)$ is defined in a similar way.  The main difference is that now $\alpha$ corresponds to a parameter in a multiple testing procedure rather than size of $\phi(X;\alpha)$.  For example, the smallest $\alpha$ allowing for $H_{0m}$ to be rejected by the Bonferroni MTP is clearly $Mp_m(x_m,u_m)$ and is referred to as the Bonferroni adjusted $p$-value.  In general, define the adjusted $p$-value by
\begin{equation}\label{adjusted p}
q_m(\mat{x},\mat{u}) = \inf\{\alpha: \delta_m(\mat{x},\mat{u};\alpha) = 1\},
\end{equation}
where $\delta_m(\mat{x},\mat{u};\alpha)$ is the decision function for $H_{0m}$ for an MTP. As before, when $\mat{u}$ is specified (generated) then $q_m(\mat{x},\mat{u})$ is referred to as an \textbf{adjusted nonrandomized (randomized) $p$-value}. When $\mat{U}$ is an unrealized random vector, then $q_m(\mat{x},\mat{U})$ is a random variable and is referred to as an \textbf{adjusted abstract randomized $p$-value} or \textbf{adjusted fuzzy $p$-value}.

\begin{figure}
\center
\epsfig{file = 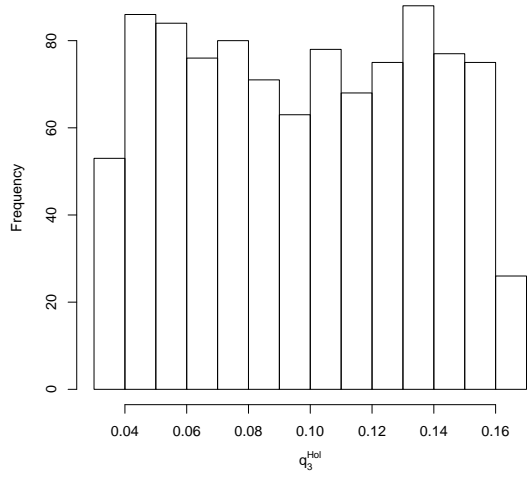, width = 3in}
\caption{\label{q-value simul} The simulated distribution of $q_3^{Hol}(\mat{x},\mat{U})$.}
\end{figure}

The distribution of the adjusted abstract randomized $p$-value can be difficult to derive analytically, but is not difficult to graph via simulation.  To illustrate, recall the example in Table \ref{illustrate}, where the goal was to test $H_{0m}:p_m = 1/2$ vs. $H_{1m}:p_m>1/2$ with data $\mat{X} = (10, 9, 8, 6, 5)$.  To estimate the distribution of a Holm-adjusted abstract randomized $p$-value we 1) generate $\mat{U}^b = \mat{u}^b$ as above and compute $p_m^*(x_m,u_m^b)$ for each $m$ 2) get the the Holm-adjusted randomized $p$-value $q_m^{Hol}(\mat{x},\mat{u}^b)$ and 3) repeat the above steps for $b=1, 2, ..., B$ and construct a histogram of $q_m^{Hol}(\mat{x},\mat{u}^1), q_m^{Hol}(\mat{x},\mat{u}^{2}), ..., q_m^{Hol}(\mat{x},\mat{u}^B)$.  A histogram of $q_3^{Hol}(\mat{x},\mat{u}^1), q_3^{Hol}(\mat{x},\mat{u}^2), ..., q_3^{Hol}(\mat{x},\mat{u}^{1000})$ is presented in Figure \ref{q-value simul}.  Each adjusted randomized $p$-value $q_m^{Hol}(\mat{x},\mat{u}^b)$ is computed using the $p.adjust$ function in R.

Recall in Table \ref{illustrate} that $\phi_3^{Hol}(\mat{x};0.05) = 0.13$ and observe it appears that about 13\% of the $q_3^{Hol}(\mat{x},\mat{u}^b)$s fall below 0.05.
We formalize this link between $\phi_m(\mat{x};\alpha)$ and $q_m(\mat{x},\mat{U})$ in Theorem \ref{theorem 3} below.
\begin{theorem} \label{theorem 3}
Let $q_m(\mat{x},\mat{u})$ be an adjusted $p$-value defined as in (\ref{adjusted p}) and suppose that $\bm{\phi}(\mat{x};\alpha)$ is a multiple test function defined as in Definition \ref{mtf}.  Further, let $\mat{U}$ be a collection of i.i.d. uniform(0,1) random variables and suppose that $\mat{U}$ and $\mat{X}$ are independent. Then for every $m$ and $\mat{x}\in\mathcal{X}$, $\Pr_{\mat{U}}(q_m(\mat{x},\mat{U})\leq\alpha) = \phi_m(\mat{x};\alpha)$.
\end{theorem}
As in Theorem \ref{theorem 1}, the above result has practical implications.  In particular, we only need to be concerned with adjusted abstract randomized $p$-values when $\phi_m(\mat{x};\alpha) \in (0,1)$.  If $\phi_m(\mat{x},\mat{u};\alpha)$ is equal to 0 or 1, then the adjusted nonrandomized and randomized $p$-value will lead to the same conclusion regardless of the value of $\mat{u}$.  Additionally, if it is computationally cumbersome to plot the distribution of $q_m(\mat{x},\mat{U})$, Theorem \ref{theorem 3} can be used to describe it.  For example, if $\phi_m(\mat{x};\alpha)>0.5$ then the median of $q_m(\mat{x},\mat{U})$ is less than $\alpha$.

\subsection{The method}
We are now in position to formally describe the method.  Again, assume that $\mat{X} = \mat{x}$ is observed.
\begin{enumerate}
\item[1.]\textit{Compute $\bm{\phi}(\mat{x};\alpha)$.  For each $\phi_m(\mat{x};\alpha) = 1(0)$ reject(retain) $H_{0m}$.  For each $\phi_m(\mat{x};\alpha)\in(0,1)$, report $\phi_m(\mat{x};\alpha)$ and $q_m(\mat{x},\mat{U})$ and go to 2a. or 2b.
\item[2a.] Generate $\mat{U} = \mat{u}$, a collection of $M$ i.i.d. uniform$(0,1)$ variates and compute $\delta_m(\mat{x},\mat{u};\alpha)$ and $q_m(\mat{x},\mat{u})$ for each relevant $m$.  If $\delta_m(\mat{x},\mat{u};\alpha) = 1(0)$ or if $q_m(\mat{x},\mat{u})\leq(>)\alpha$ then reject(retain) $H_{0m}$ for each relevant $m$.
\item[2b.] Specify $\mat{u}$ and compute $\delta_m(\mat{x},\mat{u};\alpha)$ and $q_m(\mat{x},\mat{u})$ for each relevant $m$.  If $\delta_m(\mat{x},\mat{u};\alpha) = 1(0)$ or if $q_m(\mat{x},\mat{u})\leq(>)\alpha$ then reject(retain) $H_{0m}$ for each relevant $m$.}
\end{enumerate}
Observe that for most null hypotheses, a decision will be made in Step 1.  However, for large $M$, it is now likely that Step 2 will need to be implemented for one or more null hypothesis, as we will see in the next section.  Hence, we are now forced to address the fact that the value of $\mat{u}$ will impact some of the final decisions, whether it is generated or specified.

\subsection{Assessment}

As in the single testing setting, we assess the performance of the method by comparing $\bm{\phi}(\mat{X};\alpha)$ to the multiple decision functions in Steps 2a and 2b.
\begin{theorem} \label{theorem 4}
Let $\bm{\phi}(\mat{x};\alpha)$ be defined as in Definition \ref{mtf} and let $\mat{U}$ be a collection of i.i.d. uniform(0,1) random variables.  Assume \mat{U} and $\mat{X}$ are independent.   Then
\begin{enumerate}
\item[C4:] $E_{\mat{U}}[\delta_m(\mat{x},\mat{U};\alpha)] = \bm{\phi}_m(\mat{x};\alpha)$ for every $m$ and $\mat{x}\in\mathcal{X}$ and hence $E_{(\mat{X},\mat{U})}[\bm{\delta}_m(\mat{X},\mat{U};\alpha)] = E_{\mat{X}}[\bm{\phi}_m(\mat{X};\alpha)]$ for every $m$. Further,
\item[C5:] $Var(\phi_m(\mat{X};\alpha))\leq Var(\delta_m(\mat{X},\mat{U};\alpha))$ for each $m$.
\end{enumerate}
Suppose $p$-values are computed $p_m^*(x_m, u_m)$ as in (\ref{example pu}) and consider a single step MTP defined via $\delta_m^*(x_m,u_m;\alpha) = I(p_m^*(x_m,u_m)\leq c\alpha )$ for some positive constant $c$.  Then for any specified $u_m\in (0,1)$
\begin{enumerate}
\item[C6:] $E_{X_m}[\delta_m^*(X_m,u_m;\alpha)]$ is nonincreasing in $u_m$ and $E_{X_m}[\delta_m^*(X_m,u_m;\alpha)]\neq E_{X_m}[\phi_m^*(X_m;\alpha)]$ for every $\gamma_m(\alpha)\in(0,1)$ with $E_{X_m}[\delta_m^*(X_m,u_m;\alpha)] >(<) E_{X_m}[\phi^*(X_m;\alpha)]$ for $u_m\leq (>)\gamma_m(\alpha)$.
\end{enumerate}
\end{theorem}
The interpretations of Claims C4 - C6 are akin to the interpretations of Claims C1 - C3, respectively.  In particular, Claims C4 and C5 state that the randomized multiple decision functions are unbiased and that information lost in reporting only $\bm{\delta}(\mat{X},\mat{U};\alpha)$ in Step 2a is manifested as variability.  Claim C6 states that the nonrandomized decision rule in Step 2b is biased and that the nature of the bias (conservative or liberal) can be understood by comparing $\phi_m(\mat{x};\alpha)$ to $u_m$.
\section{Illustrations}
We consider two examples.  The first deals with an analysis of microarray data while the second considers multiple hypothesis tests about a binomial proportion.  In the first example all test statistics have identical supports, but due to the complexity of the multiple testing procedure the distributions of the adjusted abstract randomized $p$-values must be estimated numerically.   In the second example, the supports of the test statistics vary and thus a multiple testing procedure which takes advantage of this property is considered.

\subsection{Microarray example}
Consider the microarray data in \cite{Tim07}, where 24 microarray chips gave rise to 10 brown fat cell gene expression measurements and 14 white fat cell gene expression measurements for each of 12,488 genes (see Table \ref{data}).  The basic goal is to compare gene expression measurements across treatment groups for each gene.
\begin{table}[h!]\begin{center}
\caption{A portion of the microarray data in \cite{Tim07}. \label{data}}
\begin{tabular}{|c|cccc|cccc|}\hline
m & $x_{m,1}$&$x_{m,2}$&...&$x_{m,10}$&$y_{m,1}$&$y_{m,2}$&...&$y_{m,14}$\\ \hline
1&1.22&1.66&...&1.41&5.64&1.79&...&11.50\\
2&3.57&19.22&...&5.23&5.17&29.49&...&7.58 \\
.&.&.&...&.&.&.&...&.\\
.&.&.&...&.&.&.&...&.\\
M=12488& 2.52 & 10.91&... & 22.67 & 10.70 & 7.35 & ... & 21.95\\  \hline
\end{tabular}\label{data}
\end{center}
\end{table}

Formally assume that $\bm{X}_m = (X_{m,1},...,X_{m,n_x}) \stackrel{i.i.d.} \sim F_m(\cdot)$ and $\bm{Y}_m = (Y_{m,1},...,Y_{m,n_y}) \stackrel{i.i.d.} \sim F_m(\cdot - \theta_m),$ where $F_m$ is a distribution function, $\theta_m \in (-\infty,\infty)$ is a location shift parameter, and $\mat{X}_m$ and $\mat{Y}_m$ are brown and white fat cell expression measurement for gene $m$, respectively. For each $m$, we test $H_{0m}:\theta_m = 0$ vs. $H_{1m}:\theta_m \neq 0$ with the two-sample Wilcoxon rank sum statistic \citep{Wil47} defined
$$
w_m = w(\bm{x}_m,\bm{y}_m) = \sum_{j=1}^{n_y} R_j(\bm{x}_m,\bm{y}_m); \textrm { } R_j(\bm{x}_m,\bm{y}_m) = \textrm{ rank of }y_{m,j}\textrm{ among }(\bm{x}_m,\bm{y}_m),
$$
Note that $E_{W_m}^0(W_m) = n_yn_x/2$ so that large values of $w_m^* = |w_m - n_yn_x/2|$ are evidence against $H_{0m}$.  Thus we compute a $p$-value via
$p_m^*(w_m^*, u_m) = \Pr^0_{W_m}(W_m^*>w_m^*) + u_m\Pr^0_{W_m}(W_m^* = w_m^*)$
for each null hypothesis as in (\ref{example pu}). Denote the collection of test statistics by $\mat{w}^* = (w_1^*, w_2^*, ..., w_M^*)$ and the collection of $p$-values by $\mat{p}^*(\mat{w}^*,\mat{u})$.

Typically the goal in a microarray analysis is to reject as many null hypotheses as possible subject to the constraint that the FDR is less than or equal to $\alpha$ for some $\alpha$, say $\alpha = 0.05$.  Here, we consider the adaptive FDR method in \cite{Sto02, StoTaySie04}, which is based on an estimator for the FDR at threshold $t$ defined by
$$ \widehat{FDR}^{\lambda}(t) = \hat{M}_0(\lambda) \frac{ t}{R(t)},$$
where $$\hat{M}_0(\lambda) = \frac{\{\# p_m^*(w_m^*,u_m)>\lambda\} + 1}{1-\lambda}$$ is an estimate of the number of true null hypotheses depending on tuning parameter $\lambda$ and $\mat{p}^*(\mat{w}^*,\mat{u})$, and where $R(t) = \{\# p_m^*(w_m^*,u_m) \leq t\}$ is the number of rejected null hypotheses if rejecting $H_{0m}$ when $p_m^*(w_m^*,u_m)\leq t$.   \cite{Sto02} defines the FDR-adjusted $p$-value for $H_{0m}$ by
$q_m^{S}(\mat{x},\mat{u}) = \widehat{FDR}^{\lambda}(p_m^*(w_m^*,u_m))$ and the decision function for each $H_{0m}$ is defined
$\delta_m^{S}(\mat{w}^*,\mat{u}) = I(q_m^{S}(\mat{w}^*,\mat{u})\leq \alpha)$.  We take $\lambda = \alpha = 0.05$ here because \cite{BlaRoq09} demonstrated that choosing $\lambda = \alpha$ ensures that the FDR is less than or equal to $\alpha$ under most dependence structures, while other choices of $\lambda$ need not lead to FDR control if data are dependent, which may be the case for this data set.  Computations are performed using the $q-value$ package in R \citep{qva}.

\begin{figure}
\center
\epsfig{file = 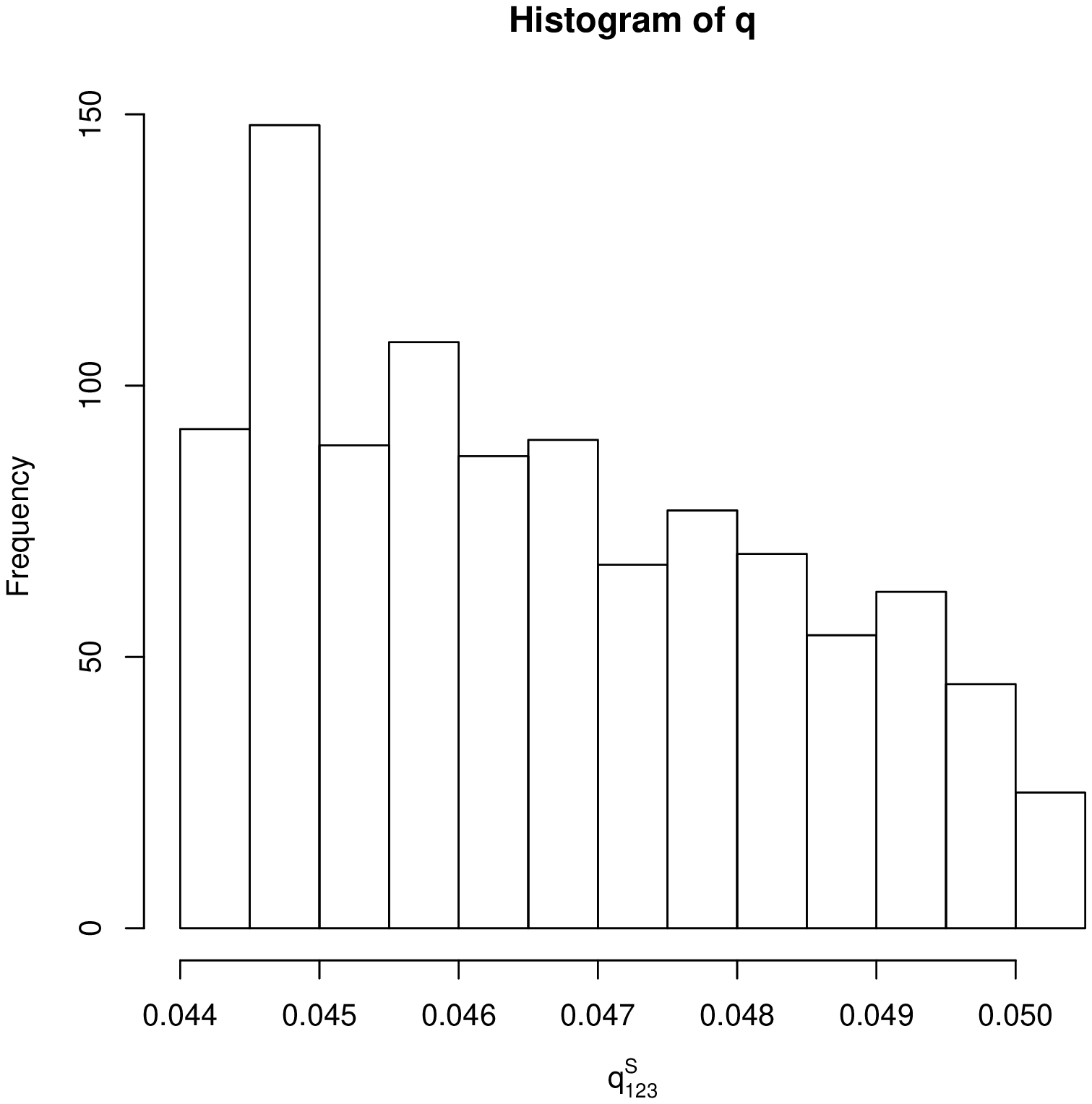, width = 3 in}
\caption{\label{fuzzy q-values} A histogram of $q_{123}(\mat{w}^*,\mat{u}^1), q_{123}(\mat{w}^*, \mat{u}^2), ..., q_{123}(\mat{w}^*, \mat{u}^B)$.}
\end{figure}

For Step 1 of the method, we generate $\mat{u}^{1}, \mat{u}^2, ..., \mat{u}^B$ vectors of i.i.d uniform$(0,1)$ variates and compute $\phi_m^{S} (\mat{w}^*;0.05) =B^{-1}\sum_{b = 1}^{B}\delta_m^{S}(\mat{w}^*,\mat{u}^{b};0.05)$ with $B=1000$ for each of the $12,488$ null hypotheses. Here, $\phi_m^S(\mat{w}^*;0.05) = 1$ for 3033 genes and $\phi_m^S(\mat{w}^*;0.05) = 0$ for 9302 genes, while $\phi_m^S(\mat{w}^*;0.05) = 0.94$ for the remaining 153 genes, which we refer to as the third group of genes hereafter.  See row 3 of Table \ref{results} entitled ``Step 1.''  The abstract adjusted $p$-value for gene $123$, which was in the third group, is presented in Figure \ref{fuzzy q-values}. The abstract adjusted $p$-values of the other genes in Group 3 had the same distribution, and hence are not presented.
\begin{table}\begin{center}
\caption{\label{results} Summary of the results of Step 1, Step 2a  and Step 2b when choosing $\mat{U} = \bm{0.5}$, $\mat{U} = \bm{1}$, and $\bm{U} = \bm{u}^*$ (see Section 5) in the analysis of the \cite{Tim07} data.}
\begin{tabular}{|c|ccc|} \hline
& \multicolumn{3}{c|}{Value of test/decision function} \\ \cline{2-4} \cline{2-4}
Procedure & $0$ & $1$ & $0.94$\\ \hline
Step 1 & $9302$ & $3033$ & $153$ \\
Step 2a (Generate $\bm{U}$) & $9315$ & $3173$ & $0$ \\
Step 2b ($\bm{U} = \bm{0.5}$) & $9302$ & $3186$ & $0$ \\
Step 2b ($\bm{U} = \bm{1}$) & 9455 & 3033&0 \\
Step 2b ($\bm{U} = \bm{u}^*$) & $9303$ & $3185$ & $0$ \\ \hline
\end{tabular}
\end{center}
\end{table}
The histogram reveals that indeed $94\%$ of the distribution of $q_{123}^S(\mat{w}^*,\mat{U})$ is below $0.05$, as Theorem \ref{theorem 3} suggests. Additionally, we see that the entire distribution  is below $0.051$ and above $0.044.$  In row 4 of Table \ref{results}, we see that the adjusted randomized $p$-values allow for $3173$ discoveries while the adjusted mid-$p$-values (see row 5) result in $3186$ discoveries.  The adjusted natural $p$-values (row 6) allowed for 3033 discoveries.  Row 7 will be discussed in Section 5.

As Theorem \ref{theorem 4} suggest, $\phi_m(\mat{w}^*;0.05)$ helps us understand the discrepancies between the above decisions in Steps 2a and 2b.  In particular, $\phi_m^S(\mat{w}^*;\alpha) = 0.94$ suggests that the randomized approach should randomly discover about 94\% of genes in Group 3.  Indeed, we see in Table \ref{results} that 140 out of the 153 were discovered when generating $\mat{U}$.  When specifying $\mat{U} = \bm{0.5}$, the mid-$p$-value approach discovers all 153 genes.  Theorem \ref{theorem 4} indicates that this approach is liberal because $u_m = 1/2 \leq 0.94$.  Indeed, it allows for 13 additional rejections over the unbiased randomized approach.  On the other hand, if opting for natural $p$-values via $\mat{U} = \bm{1}$, none of the 153 null hypotheses from Group 3 are rejected. This approach is conservative because $u_m = 1 >0.94$.

\subsection{Binomial example}
The fact that each test function took on values $0$, $1$ or $0.94$ in the previous example is attributable to the fact the support of each test statistic is identical.  However, in many applications the supports of the test statistics will vary, and improvements over traditional multiple testing procedures can be made by automatically retaining a null hypothesis whenever the support of the test statistic is highly discrete and applying multiplicity adjustments to the remaining hypotheses.  See, for example, \cite{Tar90, WesWol97, Rot99, Gil05, GutHoc07}.

To illustrate this approach and how the methodology outlined in Section 3 can be applied, $X_m$ was generated from a binomial distribution with success probability $p_m = 0.5$ for $m = 1, 2, ..., 25$ and $p_m = 0.8$ for $m = 26, 27, ..., 50$.  Sample sizes $n_1, n_2, ..., n_{50}$ were generated from a Poisson distribution with mean $\mu = 15$. Let us suppose that the goal is to test $H_{0m}:p_m = 1/2$ against $H_{1m}:p_m>1/2$ for $m = 1, 2, ..., M = 50$.  A portion of simulated data is in Table \ref{Tarone procedure}, columns 2 and 3.

We consider the modified Bonferroni MTP in \cite{Tar90} with $\alpha = 0.05$, which utilizes natural $p$-values computed $p_m^*(x_m,1) = \Pr_{X_m}^0(X_m\geq x_m)$.  The idea is to automatically retain all null hypotheses whose smallest possible natural $p$-value $p_m^*(n_m,1) = \Pr_{X_m}^0(X_n = n_x)$ would not allow for $H_{0m}$ to be rejected with the usual Bonferroni procedure.  We shall refer to this step as Step 0 because it could be applied before data collection.  Then, the procedure applies the Bonferroni multiplicity adjustments to the remaining natural $p$-values.  This step is referred to as Step 2b to be consistent with our method in Section 3.  For example, in Column 4 of Table \ref{Tarone procedure} we see that $p_m^*(n_m,1)>0.05/50$ for $m = 1, 2, 3$ while $p_m^*(n_m,1)\leq 0.05/47$ for $m = 4, 5, ..., 50$.  Hence, the Step 0 automatically retains $H_{0m}$ for $m = 1, 2, 3$ and then Step 2b rejects $H_{0m}$ if $q_m^{Tar}(x_m,1) = 47p_m^*(x_m,1) \leq 0.05$ for the remaining 47 null hypotheses.
\begin{table}[h!]
\begin{center}
\caption{\label{Tarone procedure} Depiction of the data, sorted by $n_m$ in ascending order}
\begin{tabular}{cccccc}
m&$x_m$& $n_m$ & $p_m(n_m,1)$ & $p_m(n_m,1/2)$   \\ \hline
1 & 6 & 8 & 0.0039& 0.0020  \\ \hline
2 & 5 & 9 & 0.0020 &0.0009  \\
3 & 6 & 9 & 0.0020 & 0.0009  \\ \hline
4 & 4 & 10 & 0.0009& 0.0005  \\
5 & 7 & 11 & 0.0005& 0.0002 \\
\vdots & \vdots& \vdots&\vdots & \vdots \\
50 & 21 & 27 & $<10^{-9}$ & $<10^{-10}$    \\
\hline
\end{tabular}
\end{center}
\end{table}

The procedure resulted in a total of 4 rejected null hypotheses and 46 retained null hypotheses.  However, had we additionally reported Step 1, we would have learned that 6 null hypotheses were not rejected due to the choice of $u_m = 1$ in Step 2b.  We again refer to this collection of null hypotheses as Group 3.  Test functions $\phi_m^{Tar}(x_m;0.05) = \phi_m^*(x;0.05/47)$ and adjusted abstract randomized $p$-values $q_m^{Tar}(x_m,U_m)$ are summarized in Table \ref{Tarone 2}.  Note that $q_m^{Tar}(x_m,U_m) = 47[\Pr_{X_m}^0(X_m>x_m) + U_m \Pr_{X_m}^0(X_m=x_m)]$ is a uniformly distributed random variable with lower limit $q_m^{Tar}(x_m,0)$ and upper limit $q_m^{Tar}(x_m, 1)$.  Hence, we may recover $\phi_m^*(x_m;0.05/47) = [0.05 - q_m^{Tar}(x_m,0)]/[q_m^{Tar}(x_m, 1) - q_m^{Tar}(x_m,0)]$ for those $m$ in Group 3 due to Theorem 3.
\begin{table}[h!]\begin{center}
\caption{\label{Tarone 2}Test functions and adjusted abstract randomized $p$-values for 6 tests in Group 3. }
\begin{tabular}{ccccc}
$x_m$ & $n_m$ & $\phi_m^{Tar}(x_m;0.05)$ & $q_m^{Tar}(x_m,0)$ & $q_m^{Tar}(x_m, 1)$ \\ \hline
11 & 12 & 0.280 & 0.011 & 0.149 \\
11 & 12 & 0.280 & 0.011 & 0.149 \\
12 & 14 & 0.027 & 0.043 & 0.304 \\
12 & 14 & 0.027 & 0.043 & 0.304 \\
19 & 23 & 0.777 & 0.011 & 0.061 \\
21 & 27 & 0.139 & 0.036 & 0.139 \\ \hline
\end{tabular}
\end{center}
\end{table}
If we apply Step 2a to the 6 null hypotheses in Group 3, we would expect to randomly reject an additional $0.280 + 0.280 + ... + 0.139 = 1.53$ null hypotheses. See column 3, Table \ref{Tarone 2}.

Observe that we could consider a modified mid-$p$-based Tarone method which automatically retains a null hypothesis if $p_m^*(n_m,1/2) = 0.5\Pr_{X_m}^0(X_m = n_x)>0.05/50$ in Step 0, implements Step 1 as before and then applies Bonferroni adjustments to the remaining mid-$p$-values in Step 2b.  This results in 1 unrejected null hypothesis in Step 0 (see Table \ref{Tarone procedure}), 4 null hypothesis being automatically rejected in Step 1, and an additional null hypothesis being rejected due to the specification of $u_m = 1/2$ in Step 2b.  Hence a total of 6 null hypotheses are rejected.

\section{Towards unbiased adjusted nonrandomized $p$-values}
A relevant question is: ``Why choose $u_m = 1/2$  or $u_m = 1$ for each $m$ in the computation of the nonrandomized $p$-values in Step 2b of the microarray analysis in Section 4.1?''  For example, had we choose some $u_m$s near 0 and some $u_m$s near 1, then we could have safeguarded against a strategy which would either reject or retain all null hypotheses in Group 3 in our microarray example. This idea is addressed in this section.

To better understand the benefits of such a strategy, recall in the discussion of Theorem 2 that if testing a single null hypothesis the choice of $u = 1/2$ is ideal in the sense that the supremum of the bias quantity $|I(u\leq \gamma(\alpha)) - \gamma(\alpha)|$, taken over $\gamma(\alpha)$, is minimized when choosing $u = 1/2$. However, when considering multiple hypotheses, such a choice is no longer ideal. Denote $\mat{u}^* = (1/[M+1], 2/[M+1], ..., M/[M+1])$, which is composed of the expected values of the order statistics of $M$ i.i.d. uniform$(0,1)$ random variables.  Consider a single-step multiple testing procedure defined by $\delta_m^*(x_m,u_m;\alpha) = I(u_m\leq \phi_m^*(x_m;\alpha))$ and suppose that $\gamma_m(\alpha) = \gamma(\alpha)$ for each $m$, as in Sections 3 and 4.1.  Consider the total bias quantity defined by
$$\mbox{Bias}(\mat{u}) = \sum_{m=1}^{M} \left[I(u_m\leq\gamma(\alpha)) - \gamma(\alpha)\right].$$
Observe that
$$
\mbox{Bias}(\mat{u}^*) = \sum_{m=1}^{M}I\left(\frac{m}{M+1}\leq \gamma(\alpha)\right) - M\gamma(\alpha) \rightarrow 0
$$
as $M\rightarrow \infty$. On the other hand, suppose we choose $u_1 = u_2 = ... = u$ for some $u\in (0,1)$.  Then
\begin{eqnarray*}
\mbox{Bias}(\bm{u}) = \sum_{m=1}^{M} I(u\leq\gamma(\alpha)) - M\gamma
\end{eqnarray*}
is $-M\gamma$ if $u > \gamma(\alpha)$ and is $M[1-\gamma(\alpha)]$ otherwise.

Above we see that the unbiased adjusted nonrandomized $p$-value, defined $q_m(\mat{x}, \mat{u}^*)$, generally dominates the adjusted randomized $p$-value in the sense that, for large $M$, both adjusted $p$-values are approximately unbiased, with the former being less variable because $\bm{u}^*$ is fixed rather than generated.  Additionally, the unbiased adjusted nonrandomized $p$-values are less biased than the adjusted mid-$p$-values.  Applying the $q$-value procedure from Section 4.1 to the microarray data via  $\delta_m^S(\mat{w}^*,\mat{u}^*;0.05)$ yields $3185$ discoveries (see the last row in Table \ref{results}). That is, these unbiased adjusted nonrandomized $p$-values reject most, but not all, null hypotheses in Group 3.  We should remark, however, though this nonrandomized $p$-value enjoys some nice mathematical properties, and though the choice of $\mat{U} = \mat{u}^*$ seems no more or less arbitrary than $\bm{U} = \bm{0.5}$, this nonrandomized approach does have a drawback.  Namely, if we permute the elements of $\bm{u}^*$, results may vary.  This is not the case for the usual adjusted mid-$p$-value because the vector $\bm{0.5}$ is permutation invariant.

\section{Concluding remarks}
This paper provided a unifying framework for nonrandomized, randomized and abstract randomized $p$-values in both the single and multiple hypothesis testing setting.  It was shown that an abstract randomized testing approach can be viewed as the first step in a hypothesis testing procedure, and that the randomized and nonrandomized approaches correspond to a second step in which a decision is made via the specification or generation of a value $u$.  We saw that the value of the (multiple) test function and the (adjusted) abstract randomized $p$-value provide useful additional information for understanding the properties of the usual randomized and nonrandomized procedures, and consequently, should be reported in settings when the value of $u$ may affect the final decision.

Recall in our definition of $\bm{\phi}(\mat{x};\alpha)$ and in Step 2a we assumed that the $U_m$s were independently generated.  However, it is possible to compute $p$-values $p_1(X_1, U), p_2(X_2, U), ..., p_M(X_M,U)$, where $U$ is a single uniform(0,1) random variable.  Though such a route warrants further study, we do not consider it here because the resulting $p$-values are dependent and consequently the resulting MTP need not be valid. For example, the MTP in Section 4.1 has only been shown, analytically, to control the FDR under the assumption that $p$-values are independent \citep{StoTaySie04}.  Of course some multiple testing procedures, such as the well-known Bonferroni procedure, do not require $p$-values to be independent.  We have focused on independently generated $u_m$s in this paper, however, because this approach ensures that the resulting MTP will remain valid as long as the $u_m$s are independently generated, whereas the single-$u$ approach may or may not result in a valid MTP. 

As mentioned in the Introduction, it was not our goal to recommend a randomized decision rule over a nonrandomized rule or vice versa.  The main goals of this paper were to 1) illustrate that, especially in multiple hypothesis testing, the choice to opt for a randomized or nonrandomized rule will have a significant impact in the final analysis and 2) provide tools for quantifying the effect.  However, it is worth noting that, while many find the the randomized decision rule impractical in the single hypothesis testing case, recent research suggests that this approach may be more practical in multiple testing.  We refer the curious reader to \cite{HabPen11} or \cite{Dic13} for details and examples.

Examples in this paper focused on settings when large values of the test statistic $X$ are evidence against $H_0$.  However, other more complicated settings are applicable.  For example, \cite{HabPen11} show that if large or small values of $X$ are evidence against $H_0$, then the randomized $p$-value can be computed
$p(X,U) = 2\min\left\{\frac{p^*(X,U)}{U}, \frac{1 - p^*(X,U)}{1 - U}\right\}$, where $p^*(X,U)$ is defined as in (\ref{example pu}).  Using Theorem \ref{theorem 1}, a size-$\alpha$ test function can be defined by $\phi(x;\alpha) = E_U[I(p(x,U)\leq \alpha]$.  A similar strategy can be applied to even more complex fuzzy $p$-values in \cite{GeyMee05}.  Additionally, methods in this paper are applicable to other multiple testing procedures not considered here, so long as adjusted $p$-values can be computed.

In our microarray example, all test statistics had the same support and $\phi_m(\mat{w}^*;0.05)$ only took on values 0, 1 or 0.94. In similar settings, the unbiased adjusted nonrandomized $p$-values in Section 5 should generally outperform the usual adjusted mid-$p$-value in terms of bias by ensuring that the same decision is not applied to every test function not taking values 0 or 1.  However, when the supports of the tests statistics are not identical (see Section 4.2), test functions may take on different values in $(0,1)$, and the implementation of the usual mid-$p$-value may be more tractable as it still need not lead to the same decision whenever the test functions are not 0 or 1.   If testing many null hypotheses in this nonidentical support setting, it may not be practical to present a histogram of an adjusted abstract randomized $p$-value whenever a test function is not 0 or 1.  However, it may be reasonable to present an upper and lower bound for each adjusted abstract randomized $p$-value as in Section 4.2.  This, along with one or two histogram examples, may be adequate to communicate the properties the mid-$p$-value or randomized $p$-value-based decisions to the practitioner.  In fact, even just knowing which test functions were not $0$ or $1$ is insightful.

Of course it is certainly not simpler to report the value of the test function in addition to the usual randomized or nonrandomized $p$-values.  The merit of doing so will ultimately depend on the level of collaboration between the statistician and the client, and certainly some may choose to simply report the usual natural, mid-, or randomized $p$-values and a corresponding list of rejected null hypotheses.  It is our hope that, if so, methods in this paper can at least help statisticians better understand which of these $p$-value is most appropriate for the particular problem at hand.

{\center\Large \textbf{Appendix: Proofs of Theorems}}

\noindent \textbf{Proof of Theorem \ref{theorem 1}:} It can be verified using the definitions of $p(x,u)$ and $\delta(x,u;\alpha)$ that $[p(x,u)\leq\alpha] = [\delta(x,u;\alpha) = 1] = [u\leq \phi(x;\alpha)]$ with probability 1.  Hence, $\Pr_U(p(x,U)\leq\alpha) = \Pr_U(U\leq\phi(x;\alpha)) = \phi(x;\alpha)$. $\square$ \\

\noindent \textbf{Proof of Theorem \ref{theorem 2}:}  Claim C1 follows from the definition of $\delta$.  For Claim C2, by the law of iterated expectation and from Claim C1
$$
Var(\delta(X,U;\alpha))
\geq  Var(E_{U}[\delta(X,U;\alpha)]) = Var(\phi(X;\alpha))
$$
As for Claim C3, clearly $\delta^*(x,u;\alpha)$ is nonincreasing in $u$ for every $x$ and hence $E_X[\delta^*(X,u;\alpha)]$ is nonincreasing in $u$.  Now, writing $\delta^*(x,u;\alpha) = I(X>k(\alpha)) + I(u\leq\gamma(\alpha))I(x = \gamma(\alpha))$, from Claim C1
\begin{eqnarray}
E_X[\phi^*(X;\alpha)] &=& E_{(X,U)}[\delta^*(X,U;\alpha)] \nonumber\\
&=& \Pr(X>k(\alpha)) + \Pr(U\leq\gamma(\alpha))\Pr(X=k(\alpha)) \nonumber \\
&=&\Pr(X>k(\alpha)) + \gamma(\alpha)\Pr(X=k(\alpha)) \nonumber \\
&\neq& \Pr(X>k(\alpha)) +I(u\leq \gamma(\alpha))\Pr(X=k(\alpha)) \label{notequal}\\
& = &E_X[\delta^*(X,u;\alpha)] \nonumber
\end{eqnarray}
for $\gamma(\alpha)\in(0,1)$. Observe that the ``$\neq$'' in (\ref{notequal}) can be replaced with a ``$>$'' if $u>\gamma(\alpha)$ and can be replaced with ``$<$'' if $u\leq \gamma(\alpha)$. $\square$ \\

\noindent \textbf{Proof of Theorem \ref{theorem 3}:} By the definitions of $q_m(\mat{x},\mat{u})$ and $\delta_m(\mat{x},\mat{u};\alpha)$ we have that $[q_m(\mat{x},\mat{u})\leq\alpha] = [\delta_m(\mat{x},\mat{u};\alpha) = 1]$ with probability 1.  But,
$\Pr_{\mat{U}}(\delta_m(\mat{x},\mat{U};\alpha) = 1) = \phi_m(\mat{x};\alpha)$ by definition, which implies
$\Pr_{\mat{U}}(q_m(\mat{x},\mat{U})\leq\alpha) = \Pr_{\mat{U}}(\delta_m(\mat{x},\mat{U};\alpha) = 1) = \phi_m(\mat{x};\alpha)$.  $\square$ \\

\noindent\textbf{Proof of Theorem \ref{theorem 4}:} The first equality in claim C4 is satisfied by definition.  For the second equality, by the law of iterated expectation, the fact that $\mat{X}$ and $\mat{U}$ are independent, and the definition of $\bm{\phi}(\mat{X};\alpha)$, we have
$E_{\mat{(X,U)}}[\bm{\delta}(\mat{X},\mat{U})] = E_{\mat{X}}\{E_{\mat{U}}[\bm{\delta}(\mat{X},\mat{U})]\} = E_{\mat{X}}[\bm{\phi}(\mat{X})]$.
The proof of Claims C5 and C6 are similar to the proofs of Claims C2 and C3.$\square$ \\



\bibliographystyle{chicagoa}
\bibliography{MultipleTesting2}
\end{document}